\documentclass[prl,twocolumn,superscriptaddress,amsmath]{revtex4}
\usepackage{bm}

\begin{document}

\title{Reply to Comment on ``Berry phase correction to electron
  density of states in solids''}

\author{Di Xiao}
\affiliation{Department of Physics, The University of Texas, Austin,
  Texas, 78712, USA}
\author{Junren Shi}
\affiliation{Institute of Physics, Chinese Academy of Sciences,
  Beijing, 100080, China}
\author{Qian Niu}
\affiliation{Department of Physics, The University of Texas, Austin,
  Texas, 78712, USA}
\maketitle

We do not think the preceding Comment by Duval \textit{et
al}.~\cite{duval} addresses the main result of our Letter~\cite{xiao}.
Our main result is that the density of quantum states in the weak
field limit is modified from the usual form of $(2\pi)^{-d}$ to the
expression of our Eq.~(3).  This is a quantum mechanical concept, and
its application shown in Eqs.~(5-11) are all based on this
understanding.  In practice, it means that in the semiclassical limit,
when replacing the sum over states to the integral over the $\bm
k$-space, one should use the properly defined density of states,
\begin{equation}
\sum_{\bm k} \rightarrow \int \frac{d^dk}{(2\pi)^d}
\Bigl(1 + \frac{e}{\hbar}\bm B \cdot \bm \Omega\Bigr) \;.
\end{equation}
On the other hand, the phase space volume in the Liouville theorem is
a classical concept, and our discussion on it was just to motivate
our main result.

The Comment also does not contradict with any of our results in
substance.  Our claim is that the Liouville theorem on the volume
conservation in the phase space of position and momentum is violated.
Here, ``momentum'' means the gauge invariant physical momentum.  By
direct calculation, we found that the Liouville theorem can be
restored if one uses a modified measure of the phase space volume.
The authors of the Comment agree with this result, but they pointed
out an alternative route for reaching the same result apparently known
in the largely mathematical and abstract field of non-canonical
Hamiltonian dynamics~\cite{morrison}.

Their only real objection is to our consideration of the ``na\"ive
definition'' of the phase space volume from the very beginning,
because ``an abstract phase space carries no natural volume element''
which ``can only be defined through a symplectic form''.  We can
understand such a point of view if one's scope is limited only to the
mathematical object of symplectic dynamics devoid of physical meaning.
However, the purpose of our Letter is to reveal a deep misconception
that has prevented the proper application of semiclassical dynamics in
solid state physics.  Our phase space is not abstract, because the
position and momentum are well defined physical quantities.  We
started our discussion with the ``na\"ive definition'' of phase space
volume element, because that is what people naturally think of the
volume element.  Our style of direct confrontation should be more
effective in clearing out the misconception than by applying an
abstract mathematical theorem.

Why is there such a misconception?  Although the original Liouville
theorem was restricted to phase space of canonical variables, our past
experience often finds that it also applies for the physical
position-momentum variables which are gauge invariant but
non-canonical.  For example~\cite{ashcroft}, in the presence of a
magnetic field, the physical momentum $\bm k$ for an electron is
related to the canonical momentum $\bm q$ by $\bm k = \bm q + e \bm
A(\bm r)$, where we have taken $\hbar=1$ and electron charge as $-e$.
For a Bloch electron, $\bm q$ is the wave vector and also called
crystal momentum.  However, the Liouville theorem applies to the phase
space of either set of variables.  This can be explained by the fact
that the Jacobian for volume transformation between the variables of
$(\bm r,\bm q)$ and $(\bm r,\bm k)$ is unity.  However, this is true
only when the Berry curvature is zero.  For another example, in the
absence of a magnetic field, the physical position $\bm r$ can be
expressed in the form $\bm r = \bm R + \bm{\mathcal A}_n(\bm k)$,
where $\bm{\mathcal A}_n(\bm k)$ is the Berry connection related to
the Berry curvature by $\bm\Omega_n(\bm k) = \bm\nabla_{\bm
k}\times\bm{\mathcal A}_n(\bm k)$.  In this case, $\bm R$ and $\bm k$
form a canonical set of variables, but the Liouville theorem applies
whether one uses $\bm r$ or $\bm R$.

In summary, the Comment addresses an important but not the main result
of our Letter, it does not contradict our results in substance, and
the only objection is really on the style of approach.  The concrete
and rich physical results Eqs.~(3-14) revealed in our Letter in fact
conform with the general theory of symplectic dynamics, which is
usually discussed in abstract setting. It is good to make connection
with the abstract theory, and we have done some, if not thoroughly,
towards the end of our Letter.

\end{document}